# A New Phase of Tethered Membranes: Tubules

Leo Radzihovsky

*Department of Physics, University of Colorado, Boulder, CO 80309*

John Toner

*IBM Research Division, IBM T. J. Watson Research Center, Yorktown Heights, NY 10598*
*Department of Physics, University of Oregon, Eugene, OR 97403-1274* *

(October 17, 1995)

We show that fluctuating tethered membranes with *any* intrinsic anisotropy unavoidably exhibit a new phase between the previously predicted "flat" and "crumpled" phases, in high spatial dimensions $d$ where the crumpled phase exists. In this new "tubule" phase, the membrane is crumpled in one direction but extended nearly straight in the other. Its average thickness is $R_G \sim L^{\nu_t}$ with $L$ the intrinsic size of the membrane. This phase is more likely to persist down to $d = 3$ than the crumpled phase. In Flory theory, the universal exponent $\nu_t = 3/4$, which we conjecture is an exact result. We study the elasticity and fluctuations of the tubule state, and the transitions into it.

64.60Fr,05.40,82.65Dp

Tethered membranes are of great interest in large part because their behavior is much richer than that of polymers, their one-dimensional analog. Specifically, polymerized membranes have been predicted [1] to undergo a "crumpling" transition between the "crumpled" and long-ranged orientationally ordered "flat" phases. This apparent violation of the Mermin-Wagner theorem is made possible by "anomalous elasticity" [1], [2]: thermal fluctuations infinitely enhance the membrane's effective bending rigidity $\kappa$, stabilizing the orientational order against these very fluctuations.

Most past theoretical work [3] has been restricted to *isotropic* membranes. Here we consider *intrinsically anisotropic* membranes and find that this seemingly innocuous modification has profound and surprising consequences: an entire new and heretofore unanticipated phase of the membrane, which we call the "tubule" phase ubiquitously intervenes between the crumpled and "flat" phases (see Fig.1). Only in the special case of perfectly isotropic membranes, which follow a path like $P_2$, is a direct crumpled-to-flat (CF) transition possible. Generic paths like $P_1$ have *two* phase transitions, crumpled-to-tubule (CT), and tubule-to-flat (TF), which we are currently studying in an $\epsilon$-expansion [4].

There are a number of possible experimental realizations of anisotropic membranes. One is polymerized membranes with in-plane tilt order. *Fluid* membranes with such order have already been found [5]; it should be possible to polymerize these without destroying the tilt order. Secondly, membranes could be fabricated by crosslinking DNA molecules trapped in a fluid membrane [5]. Performing crosslinking in an applied electric field would align the DNA and "freeze in" the anisotropy induced by the electric field, which could then be removed.

Simulations could be done on, e.g., triangular or rectangular nets of balls and springs with all of the spring constants in one direction different, by a factor of order 2 or so, from those in the other direction. Equivalently, one could have different bond *lengths* in the two directions, or use second nearest neighbor springs of different strengths to create different bend stiffnesses in the two directions. *Any* such modification *whatsoever* will lead, upon renormalization, to a membrane with *all* of the anisotropic terms we consider here, and, hence, will fall into the universality class of our model.

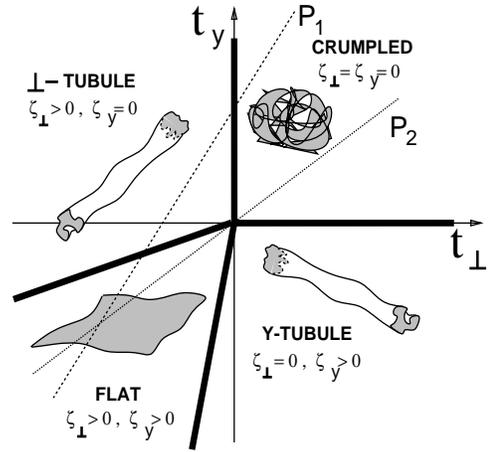

FIG. 1. Phase diagram for anisotropic tethered membranes showing the flat, crumpled and new tubule phases.

The defining property of the tubule phase is that the membrane is crumpled in one direction (y) , but "flat" in the other. Its average shape is a long, thin cylinder of length $R_y = L_y^o \times O(1)$ and radius $R_G(L_\perp) \ll L_\perp$, where $L_y$ and $L_\perp$ are the intrinsic dimensions of the membrane. The tubule radius $R_G$, and its undulations $h_{rms}$ transverse to the y-axis, obey the scaling laws:

$$R_G(L_\perp, L_y) \propto L_\perp^{\nu_t} \, , \qquad (1)$$

$$h_{rms}(L_\perp, L_y) = L_y^\zeta f_h((\Lambda L_y)/(\Lambda L_\perp)^z) , \quad (2)$$

where the universal exponents $z = \frac{1+\nu_t}{2}$, $\zeta = \frac{1-\nu_t}{2(1+\nu_t)}$ are $< 1$, $\Lambda$ is an ultraviolet cutoff, $f_h(u) \to constant$ for $u \to 0$ and $f_h(u) \propto u^{\frac{3}{2}-\zeta}$, for $u \to \infty$.

For general spatial dimension $d$, Flory theory treatment of the self-avoidance (SA) predicts $\nu_t = \frac{3}{d+1}$, suggesting that the tubule phase should be stable down to the lower critical dimension $d_{lc} = 2$, where $\nu = 1$, and therefore should exist in 3d, predicting $\nu_t = \frac{3}{4}$, which implies that $z = \frac{7}{8}$, and $\zeta = \frac{1}{14}$. However, the analogous Flory result for the crumpled phase $\nu = \frac{4}{d+2}$ has more recently been called into question. Numerical simulations [6] find no crumpled phase below $d = 4$. An uncontrolled Gaussian approximation [7,8] supports this finding, predicting $\nu = \frac{4}{d}$, which suggests that $d_{lc} = 4$ for the crumpled phase. Both this, and the numerical values of $\nu$ for $d > 4$, agree well with the simulations.

The same approximation for the tubule phase [4] gives $\nu = \frac{7}{3d-5}$, which, unfortunately, suggests that $d_{lc} = 4$ for the tubule phase as well. However, despite its success for the crumpled phase, this Gaussian approximation is known to be far from trustworthy. For example, it predicts $\nu = \frac{2}{d}$ for linear polymers, which not only is less accurate than the Flory result $\nu = \frac{3}{d+2}$, but also *incorrectly* predicts $d_{lc} = 2$ for polymers, when, in fact, it is well known that $d_{lc} = 1$ in that case.

Whether the Gaussian approximation is any more reliable for our tubule phase remains an open question. One could argue that a slice perpendicular to the y-axis through our tubule looks like a SA random walk in two dimensions, for which the Flory result of $\nu = 3/4$ is known to be *exact*, while the Gaussian approximation $\nu = 1$ is clearly wrong. Whether or not this analogy holds, it *is* clear that SA, though a relevant perturbation, is much less important for tubules than for the crumpled phase, since points on the membrane widely seperated in the y-direction never bump into each other in the tubule phase, while they do in the crumpled phase. So it seems quite plausible, the Gaussian approximation notwithstanding, that the tubule phase *is* stable in $d = 3$. Furthermore, the suppression of the crumpled phase by SA, makes the possibility of the new tubule phase even more interesting and important. Whether the tubule phase *does* survive in $d = 3$ can only be determined by simulations and experiments on anisotropic membranes, both of which we hope our work stimulates.

In the following discussions, numerical estimates for the values of the exponents will be obtained from the Flory estimate $\nu_t = \frac{3}{4}$ in 3d; these numbers should be taken with a grain of salt, due to the uncertainties just discussed, about the validity of the Flory theory.

Eq.2 implies that $h_{rms}(L) \propto L^{1-\nu_t} \approx L^{1/4}$, for a square membrane with $L_\perp \sim L_y \equiv L$, using $\lim_{L\to\infty} u \equiv \Lambda L_y/(\Lambda L_\perp)^z \to \infty$. In the "linear polymer" limit $L_y \gg L_\perp$, $h_{rms} \propto L_y^{3/2}/L_\perp^{z(3/2-\zeta)} = L_y^{3/2}/L_\perp^{\nu_t+1/2} \approx L_y^{3/2}/L_\perp^{5/4}$. Defining $L_P$ to be the value of $L_y$ at which $h_{rms} = L_y$, we obtain the orientational persistence length $L_P \propto L_\perp^{2\nu_t+1} \approx L_\perp^{5/2}$. For any roughly square membrane, $L_y$ is much less than $L_P$, hence the tubule phase is stable against thermal fluctuations as $L \to \infty$.

Like the flat phase [1,2,9], the tubule phase exhibits anomalous elasticity; however, as discussed above, SA is a strongly relevant perturbation in this new phase. The tubule, swelled by the SA interaction, acts for $L_y \gg L_y^c \equiv \Lambda^{-1}(\Lambda L_\perp)^z$ like a polymer with bending rigidity $\kappa_p(L_\perp) \propto L_\perp R_G^2 \propto L_\perp^{1+2\nu_t} \approx L_\perp^{5/2}$.

Our model for anisotropic membranes is a generalization of the isotropic model [11]. We characterize the configuration of the membrane by the position $\vec{r}(\mathbf{x})$ in the $d$-dimensional embedding space of the point in the membrane labelled by a $D$-dimensional internal co-ordinate $\mathbf{x}$. In the physical case, $D = 2$ and $d = 3$. The Landau free energy $F$ is an expansion in the local tangent vectors $\partial_\alpha \vec{r}(\mathbf{x})$, keeping only the leading terms consistent with global translation and rotation invariance:

$$\begin{aligned}F[\vec{r}(\mathbf{x})] = \frac{1}{2}\int d^{D-1}x_\perp dy &\left[\kappa_\perp \left(\partial_\perp^2 \vec{r}\right)^2 + \kappa_y \left(\partial_y^2 \vec{r}\right)^2\right.\\&+ \kappa_{\perp y} \partial_y^2 \vec{r} \cdot \partial_\perp^2 \vec{r} + t_\perp \left(\partial_\alpha^\perp \vec{r}\right)^2 + t_y \left(\partial_y \vec{r}\right)^2\\&+ \frac{u_{\perp\perp}}{2}\left(\partial_\alpha^\perp \vec{r} \cdot \partial_\beta^\perp \vec{r}\right)^2 + \frac{u_{yy}}{2}\left(\partial_y \vec{r} \cdot \partial_y \vec{r}\right)^2 + u_{\perp y}\left(\partial_\alpha^\perp \vec{r} \cdot \partial_y \vec{r}\right)^2\\&\left.+ \frac{v_{\perp\perp}}{2}\left(\partial_\alpha^\perp \vec{r} \cdot \partial_\alpha^\perp \vec{r}\right)^2 + v_{\perp y}\left(\partial_\alpha^\perp \vec{r}\right)^2\left(\partial_y \vec{r}\right)^2\right]\\&+ \frac{b}{2}\int d^D x \int d^D x' \, \delta^{(d)}\left(\vec{r}(\mathbf{x}) - \vec{r}(\mathbf{x}')\right), \quad (3)\end{aligned}$$

where the $\kappa$'s, $t$'s, $u$'s, $v$'s are elastic constants, and $b$ is the SA interaction strength. The first three ($\kappa$) terms in $F$ represent the anisotropic bending energy of the membrane. The elastic constants $t_\perp$ and $t_y$ are $> 0$ at high temperatures and $< 0$ at low temperatures. When both are positive, the membrane crumples. When either is negative, the membrane extends in the associated direction, and the $u$ and $v$ quartic terms are then needed for stability. Eq.3 reduces to the model for isotropic membranes [11] when $t_\perp = t_y$, $\kappa_{\perp\perp} = \kappa_y$, $\kappa_{\perp y} = 0$, $u_{yy} = 4(\tilde{v}+u)$, $u_{\perp\perp} = u_{\perp y} = 4u$, and $v_{\perp\perp} = v_{\perp y} = 4\tilde{v}$.

In mean-field theory, we seek a configuration $\vec{r}(\mathbf{x})$ that minimizes the free energy Eq.3. Since the curvature energies $\kappa_\perp \left(\partial_\perp^2 \vec{r}\right)^2$ and $\kappa_y \left(\partial_y^2 \vec{r}\right)^2$ vanish when $\vec{r}(\mathbf{x})$ is linear in $\mathbf{x}$, we seek these minima by inserting the ansatz $\vec{r}_o^f(\mathbf{x}) = (\zeta_\perp \mathbf{x}_\perp, \zeta_y y, 0, 0, \ldots, 0)$ into Eq.3. For now neglecting the SA interaction,

$$\begin{aligned}F = \frac{1}{2}L_\perp^{D-1}L_y &\left[t_y \zeta_y^2 + t_\perp(D-1)\zeta_\perp^2\right.\\&\left.+ \frac{1}{2}u'_{\perp\perp}(D-1)^2\zeta_\perp^4 + \frac{1}{2}u_{yy}\zeta_y^4 + v_{\perp y}(D-1)\zeta_\perp^2\zeta_y^2\right]. \quad (4)\end{aligned}$$

where $u'_{\perp\perp} \equiv v_{\perp\perp} + u_{\perp\perp}/(D-1)$. Minimizing $F$ over $\zeta_\perp$ and $\zeta_y$ yields two possible phase diagram topologies. For $u'_{\perp\perp} u_{yy} > v_{\perp y}^2$, we obtain Fig.1. Both $\zeta_\perp$ and $\zeta_y$ vanish for $t_\perp$, $t_y > 0$. This is the crumpled phase: the entire membrane, in mean-field theory, collapses into the origin, $\zeta_\perp = \zeta_y = 0$ i.e., $\vec{r}(\mathbf{x}) = 0$ for all $\mathbf{x}$. In our new y-tubule phase, characterized by $\zeta_\perp = 0$ and $\zeta_y = \sqrt{|t_y|/u_{yy}} > 0$: the membrane is extended in the y-direction but crumpled in the $\perp$ directions. The $\perp$-tubule phase is the analogous phase with the $y$ and $\perp$ directions reversed. The tubule-flat boundary slopes are $u_{yy}/v_{\perp y}$ and $v_{\perp y}/u'_{\perp\perp}$ respectively. In the flat phase, both $\zeta_\perp$ and $\zeta_y \neq 0$. For $u'_{\perp\perp} u_{yy} < v_{\perp y}^2$, the flat phase disappears, and is replaced by a direct first-order transition from $\perp$-tubule to y-tubule along the locus $t_y = (v_{\perp y}/u'_{\perp\perp}) t_\perp$.

The flat and the crumpled phases of anisotropic membranes in Fig.1 are in the same universality class [10] as those of isotropic membranes [1]. In the crumpled phase, $t_\perp$, $t_y > 0$, and all other local terms in Eq.3 are irrelevant at long wavelengths. A change of variables $\mathbf{x}_\perp = \mathbf{x}' \sqrt{t_\perp/t_y}$ makes the remaining energy isotropic.

We now consider the effects of fluctuations, ignoring SA (i.e., the "phantom" membrane). Consider the y-tubule phase. To treat fluctuations, we perturb around the mean-field solution by writing $\vec{r}(\mathbf{x}) = (\zeta_y y + u(\mathbf{x}), \vec{h}(\mathbf{x}))$, where $\vec{h}(\mathbf{x})$ is a $d-1$-component vector orthogonal to $y$. Inserting the above expression for $\vec{r}$ into Eq.3 and keeping only relevant terms gives $F_{tot} = F_{mft} + F_{el}$, where $F_{mft} = \frac{1}{2} L_\perp^{D-1} L_y [t_y \zeta_y^2 + \frac{1}{4} u_{yy} \zeta_y^4]$,

$$F_{el} = \frac{1}{2} \int d^{D-1} x_\perp dy \left[ \kappa (\partial_y^2 \vec{h})^2 + t(\partial_\alpha^\perp \vec{h})^2 + g_\perp (\partial_\alpha^\perp u)^2 \right.$$
$$\left. + g_y \left(\partial_y u + \frac{1}{2}(\partial_y \vec{h})^2\right)^2 \right] , \quad (5)$$

$\kappa \equiv \kappa_y$, $t \equiv t_\perp + v_{\perp y} \zeta_y^2$, $g_y \equiv u_{yy} \zeta_y^2/2$, and $g_\perp \equiv t + u_{\perp y} \zeta_y^2$.

Note that the ratios of the coefficients of the quadratic $(\partial_y u)^2$ and the anharmonic $\partial_y u (\partial_y \vec{h})^2$ and $(\partial_y \vec{h})^4$ terms in $F_{el}$ must be $exactly$ $4:4:1$, since they must appear together as a result of expanding the rotationally invariant combination $(\partial_y u + \frac{1}{2}(\partial_y \vec{h})^2)^2$. This ratio allows us to calculate $exactly$ the long-wavelength anomalous elasticity of "phantom" tubules, as we will show in a moment.

The propagators can be read off from Eq.5, giving $\langle h_i(\mathbf{q}) h_j(-\mathbf{q}) \rangle = k_B T \delta_{ij}^\perp G_h(\mathbf{q})$, $\langle u(\mathbf{q}) u(-\mathbf{q}) \rangle = k_B T G_u(\mathbf{q})$, where $G_h^{-1}(\mathbf{q}) = t q_\perp^2 + \kappa q_y^4$, $G_u^{-1}(\mathbf{q}) = g_\perp q_\perp^2 + g_y q_y^2$, and $\delta_{ij}^\perp$ is a Kronecker delta when both indices $i$ and $j \neq y$, and is zero if either $i$ or $j = y$. The rms fluctuations in the harmonic approximation are:

$$\langle |\vec{h}(\mathbf{x})|^2 \rangle \propto \int_{L_\perp} \frac{d^{D-1} q_\perp dq_y}{(2\pi)^D} \frac{1}{t q_\perp^2 + \kappa q_y^4} \propto L_\perp^{5/2-D}, \quad (6)$$

clearly revealing that the upper critical dimension is $D_{uc} = 5/2$. Below $D_{uc}$, we expect anomalous elasticity.

However, this anomaly is $not$ manifested in the fluctuations of $\vec{h}$ alone. We can see this by integrating out the phonons $u$ exactly, the only remaining anharmonicity in the effective elastic free energy for $\vec{h}$ alone is,

$$F_{anh}[\vec{h}] = \frac{1}{4} \int_{\mathbf{k_1},\mathbf{k_2},\mathbf{k_3}} \left( \vec{h}(\mathbf{k}_1) \cdot \vec{h}(\mathbf{k}_2) \right) \left( \vec{h}(\mathbf{k}_3) \cdot \vec{h}(\mathbf{k}_4) \right) \times$$
$$\times (\mathbf{k}_1 \cdot \mathbf{k}_2)(\mathbf{k}_3 \cdot \mathbf{k}_4) V_h(\mathbf{q}) , \quad (7)$$

where $\mathbf{q} = \mathbf{k}_1 + \mathbf{k}_2$ and $\mathbf{k}_1 + \mathbf{k}_2 + \mathbf{k}_3 + \mathbf{k}_4 = 0$. The effective vertex is $V_h(\mathbf{q}) = g_y g_\perp q_y^2/(g_y q_y^2 + g_\perp q_\perp^2)$, and is irrelevant for $D > 3/2$, as can be seen by simple power counting. Thus, in $D = 2$, the elastic constants $t$, $g_\perp$ and $\kappa_y$ are finite and non-zero as $q_y \to 0$.

However, $g_y$ $is$ driven to zero as $q_y \to 0$. In a self-consistent one-loop perturbative calculation, similar to that successfully used to compute the anomalous elasticity in the flat phase [9], we find

$$g_y(\mathbf{q}) = g_y^o - \quad (8)$$
$$\int \frac{(k_B T)^3 g_y^2(\mathbf{q}) p_y^2 (p_y - q_y)^2 \, d^{D-1} p_\perp dp_y/(2\pi)^D}{(t p_\perp^2 + \kappa(\vec{p}) p_y^4)(t|\vec{p}_\perp - \mathbf{q}_\perp|^2 + \kappa(|\vec{p} - \vec{q}|)(p_y - q_y)^4)} ,$$

where $g_y^o$ is the "bare" value of $g_y$. The above argument shows that $\kappa(\vec{p})$ can be replaced by a constant, since the $\vec{h}$ elasticity is not anomalous. This self-consistent equation can be solved by the ansatz, $g_y(\mathbf{q}) = q_y^{\eta_u} f_g(q_y/q_\perp^z)$. with $z = \frac{1}{2}$, $\eta_u = 5 - 2D$, which we have verified works to $all$ orders in perturbation theory.

We now compute the $phantom$ tubule diameter $R_G$ and transverse wandering roughness $h_{rms}$, defined by $R_G^2 \equiv \langle |\vec{h}(\mathbf{L}_\perp, y) - \vec{h}(0_\perp, y)|^2 \rangle$, $h_{rms}^2 \equiv \langle |\vec{h}(\mathbf{x}_\perp, L_y) - \vec{h}(\mathbf{x}_\perp, 0)|^2 \rangle$. Because $h_{rms}$ and $R_G$ receive large contributions from $\mathbf{q}_\perp = 0$ and $q_y = 0$ zero modes, respectively, $R_G$ and $h_{rms}$, surprisingly, scale in different ways with the membrane dimensions $L_\perp$ and $L_y$. Taking into account the zero modes, we calculate $h_{rms}$ and $R_G$ by equipartition, and find the forms Eqs.1,2, with $z = 1/2$, $\nu_t = (5 - 2D)/4$, and $\zeta = 2\nu_t$. For a nearly square membrane $L_\perp \sim L_y \sim L \to \infty$, for which $\Lambda L_y >> (\Lambda L_\perp)^z$, we obtain $h_{rms} \propto L_y^3/L_\perp^{(D-1)/2} \propto L^{2-D/2}$. Thus for a $D = 2$ phantom tubule, $h_{rms} \propto L$. Unlike the flat phase, no $\log(L/a)$ correction arises, so the $(D = 2)$ $phantom$ tubule is just marginally stable, but with wild transverse undulations. These are greatly suppressed by SA, to the effects of which we now turn.

We begin by estimating the radius of gyration using Flory theory. Specializing henceforth to the physical case $D = 2$ for simplicity, we estimate the SA energy $E_{SA}$ as $E_{SA} \approx bV\rho^2$, where the volume $V$ in the embedding space occupied by the tubule is $V \approx R_G^{d-1} L_y$, and the density $\rho$ in the embedding space of the tubule is $\rho \approx M/V \propto L_\perp L_y/V$, where $M$ is the mass of the membrane. Putting these together gives $E_{SA} \propto L_y L_\perp^2/R_G^{d-1}$.

Inserting our earlier, phantom membrane result $R_G \propto L_\perp^{1/4}$ (for $D = 2$) and taking $L_\perp \propto L_y^2$, as required by anisotropic scaling, we find that $E_{SA} \propto L_y^{5-(d-1)/2}$, which goes to infinity as $L_y \to \infty$ for $d < 11$. Thus, SA is strongly relevant, and changes the long wavelength behavior of the membrane, for $d < 11$.

We can calculate $R_G$ for $d < 11$ by combining the above estimate of $E_{SA}$ with a similar scaling estimate of the elastic energy, yielding:

$$E_{FL} = \left[t_y\zeta_y^2 + u_{yy}\zeta_y^4 + t\left(\frac{R_G}{L_\perp}\right)^2\right]L_\perp L_y + b\frac{L_y L_\perp^2}{\zeta_y R_G^{d-1}} \ . \quad (9)$$

Minimizing this over $R_G$, we obtain $R_G(L_\perp) \propto L_\perp^{\nu_t}$, with $\nu_t = \frac{3}{d+1}$. For the physical case $d = 3$, this gives $\nu_t = 3/4$. Since a slice through the tubule traces out a crumpled polymer embedded in 2d, we conjecture that $\nu_t = 3/4$ is an *exact* result for the tubule thickness, since it is for 2d polymers. For a square membrane, $L_y \sim L_\perp$, it is straightforward to argue that the $q_y = 0$ zero modes do not contribute to $R_G$, and $L_\perp$ is the dominant infra–red cutoff. Hence Eq.1 gives the correct radius of gyration.

This highly nontrivial *ground state* for the *SA* tubule is *not* modified by thermal fluctuations. That is, even at $T > 0$, the variation of $R_G$ with $L_\perp$ is completely dominated by the SA energy, i.e. determined by a $T = 0$ fixed point. This can be seen by evaluating the elastic (or SA) energy with $R_G(L_\perp)$ given by this SA ground state. For an $L \times L$, membrane, this energy $E_{el} \propto L^\theta$, $\theta = \frac{6}{d+1} = \frac{3}{2}$ (for $d = 3$), is much larger than $k_B T$ as $L \to \infty$.

Using a generalization of Landau's derivation of shell theory we now calculate anomalous elasticity in the presence of SA. Bending the tubule with radius of curvature $R_c \gg R_G$, induces a strain $\epsilon \sim R_G/R_c$, which costs an additional elastic energy density $g_y(L_y, L_\perp)\epsilon^2 = g_y(L_y, L_\perp)(R_G(L_y)/R_c)^2$. Interpreting this additional energy as an effective bending energy density $\kappa_y(L_\perp, L_y)/R_c^2$ leads to the *effective* bend modulus $\kappa_y(L_\perp, L_y)$,

$$\kappa_y(L_\perp, L_y) \sim g_y(L_\perp, L_y) R_G(L_\perp, L_y)^2 \ , \quad (10)$$

Inserting $\kappa_y(L_\perp, L_y) = L_y^\eta f_\kappa(\Lambda L_y/(\Lambda L_\perp)^z)$, $g_y(L_\perp, L_y) = L_y^{-\eta_u} f_g(\Lambda L_y/(\Lambda L_\perp)^z)$ and $R_G(L_\perp, L_y) = L_y^{\nu_t} f_R(\Lambda L_y/(\Lambda L_\perp)^z)$ into above expression, we obtain a relation between the exponents $2\nu_t = z(\eta + \eta_u)$, which is satisfied by our earlier results for the phantom tubule.

The physical *SA* tubule at $T = 0$ is absolutely straight (i.e. $\zeta_y = 1$ *exactly*). This implies that the tubule stretching elastic constant $g_y = g_y^0$, its bare value, since there are neither fluctuations nor SA effects to renormalize it (in contrast to $\kappa$). Therefore $\eta_u = 0$ at $T = 0$; Eq. 10 then implies that $\kappa_o$ for the $T = 0$ SA tubule is already anomalous, and given by $\kappa_o(\mathbf{q}) = q_y^{-2\nu_t/z} f_\kappa(q_y \Lambda^z/(\Lambda q_\perp^z))$.

The effective free energy describing thermal fluctuations about this nontrivial, SA ground state is

$$F_{eff} \approx \frac{1}{2} \int \frac{dq_\perp dq_y}{(2\pi)^2} \left(\kappa_o(\mathbf{q}) q_y^4 + t q_\perp^2\right) |\vec{h}(\mathbf{q})|^2 \ . \quad (11)$$

Balancing the $\kappa q_y^4$ term in this expression with the $t q_\perp^2$ term gives the anisotropy of scaling exponent $z$, defined by $q_y \propto q_\perp^z$. We thereby obtain $z = \frac{1}{2-\nu_t/z}$, which gives $z = \frac{1+\nu_t}{2}$. Now calculating the fluctuation corrections to $g_y$ from Eq.8 using the wavevector dependent $\kappa$ found above, we find that the integral in Eq.8 converges in the infra-red and $g_y$ is finite as $\mathbf{q} \to 0$ provided $2 > \frac{5}{2}(1 - \nu_t)$. If $\nu_t$ is anywhere near its Flory value $\nu_t = 3/4$ in $D = 2, d = 3$, this condition is clearly satisfied. Thus, unlike the phantom, the SA tubule has $\eta_u = 0$, even at $T > 0$. Using this fact in Eq.10 and taking the Flory expression for $\nu_t$ in $d = 3$, we obtain, $z = \frac{1+\nu_t}{2} = \frac{7}{8}$, $\eta = \frac{4\nu_t}{1+\nu_t} = \frac{12}{7}$, $\eta_u = 0$.

Using Eq.11 to compute the transverse tubule undulations $h_{rms}$, we find $h_{rms} = L_y^\zeta f_h(\Lambda L_y/(\Lambda L_\perp)^z)$, where $\zeta = \frac{3}{2} - \frac{1+2\nu_t}{2z}$, and $f_h(x) \approx x^{(1+2\nu_t)/2z}$ and $f_h(x) \approx$ constant for $x \to \infty$ and $x \to 0$, respectively. For a roughly square $L \times L$ membrane, we are always in the $x \to \infty$ limit of the crossover function $f_h(x)$, which implies that the rms transverse undulations of the tubule are given by $h_{rms} \sim L^{3/2-(1+2\nu_t)/2} \sim L^{1/4}$. Similar arguments applied to $R_G$ yield Eq.1.

We [4] are currently investigating the effects of SA on the tubule phase in $d = d_{uc}^{SA} - \epsilon$, following the work of Ref. [12] for the crumpled phase. The scaling theory of the tubule to flat and crumpled to tubule transitions, along with a Flory theory for these transitions, will be presented in a future publication [4].

We both thank The Institute for Theoretical Physics at UCSB, where this work was initiated, for their hospitality and support under NSF grant No. PHY94-07194.
* Address after Sept. 16, 1995.